# Large second harmonic generation enhancement in SiN waveguides by all-optically induced quasi phase matching

Adrien Billat[1†], Davide Grassani[1†], Martin H. P. Pfeiffer[2], Svyatoslav Kharitonov[1], Tobias J. Kippenberg[2], Camille-Sophie Brès[1*]

[1]*Ecole Polytechnique Fédérale de Lausanne, 1015 Lausanne, Switzerland, Photonic Systems Laboratory (PHOSL), STI-IEL, Station 11, CH-1015 Lausanne, Switzerland;*

[2]*Ecole Polytechnique Fédérale de Lausanne, 1015 Lausanne, Switzerland, Laboratory of Photonics and Quantum Measurements (LPQM), SB-IPHYS, Station 3, CH-1015 Lausanne, Switzerland;*

*\*Corresponding author: Phone  +41 21 69 37866; Fax +41 21 69 31037;*

*Email: camille.bres@epfl.ch*

[†]*These authors contributed equally*

**Integrated waveguides exhibiting efficient second-order nonlinearities are crucial to obtain compact and low power optical signal processing devices. Silicon nitride (SiN) has shown second harmonic generation (SHG) capabilities in resonant structures[1] and single-pass devices[2] leveraging intermodal phase matching, which is defined by waveguide design. Lithium niobate allows compensating for the phase mismatch using periodically poled waveguides, however the latter are not reconfigurable and remain difficult to integrate with SiN or silicon (Si) circuits[3]. Here we show the all-optical enhancement of SHG in SiN waveguides by more than 30 dB. We demonstrate that a Watt-level laser causes a periodic modification of the waveguide second-order susceptibility. The resulting $\chi^{(2)}$ grating has a periodicity allowing for quasi phase matching (QPM) between the pump and SH mode. Moreover, changing the pump wavelength or polarization updates the period, relaxing phase**

**matching constraints imposed by the waveguide geometry. We show that the grating is long term inscribed in the waveguides, and we estimate a χ$^{(2)}$ of the order of 0.3 pm/V, while a maximum conversion efficiency (CE) of 1.8·10$^{-6}$ W$^{-1}$cm$^{-2}$ is reached.**

Over the last decade, integrated photonics allowed the demonstration of micrometer-scale and low power optical nonlinear devices. In particular, CMOS compatible materials as Si and SiN are the most promising for nonlinear optical signal processing based on third-order processes[4,5]. However, integrated waveguides showing significant second-order optical nonlinearity are key to enable a new range of applications. This can be self-referencing of chip-based frequency combs[6], or telecom signal up-conversion in order to perform detection directly with group-IV photodiodes[7]. On-chip frequency down-conversion for quantum optics is also an option, as second-order processes make the pump rejection easy in photon-pair generation experiments[8].

SiN exhibits a very large transparency window, from ultraviolet to mid-infrared, and a moderate second-order nonlinearity due to interface symmetry breaking and higher multipole bulk terms[9-11]. To enhance SHG in SiN, researchers used resonant structures like microresonators[1] and waveguide gratings[12], at the expenses of a reduced bandwidth. Moreover, momentum conservation for SHG in integrated waveguides is generally achieved through intermodal phase matching, implying that the effective index of the pump mode is equal to the second harmonic (SH) one, i.e. $n_p = n_{sh}$. Phase matching is therefore constrained by waveguide design.

In this letter we demonstrate the growth of the SH signal over time in a SiN waveguide pumped by a pulsed laser in the communication band (designated as the pump). Subsequent to the SH growth, by probing the waveguide with a tunable continuous wave (CW) laser (designated as the probe), we observe a clear phase matching peak centered on the pumping wavelength. Shifting the pump to another arbitrary wavelength in the 1534 nm – 1550 nm range, we notice a similar SH growth as well as a subsequent phase-matching peak at the new wavelength. Moreover, the waveguide features the same SHG level when probed over a few days. This demonstrates a persistent change in the phase-matching condition caused by the pump, as sketched in Fig. 1a. To our knowledge, this is the first demonstration of an optically induced and dynamically reconfigurable SHG enhancement in any integrated photonic platform.

The waveguides were fabricated according to the photonic Damascene process[13], which guarantees a crack-free nitride layer deposition and void-free structures of high aspect ratio. The devices are made of stoichiometric SiN buried in $SiO_2$ and exhibit a very low attenuation (0.2 dB/cm) that results in a long effective length. They are folded in meanders and have trapezoidal cross sections (see Fig. 1b). We obtained results in two different waveguides: waveguide (i) is 4 cm long and 1.5 µm wide, while waveguide (ii) is 5.8 cm long and 1.4 µm wide. Both are 0.87 µm thick and terminated by inverse taper mode converter. Fig. 1c shows the optical setup, which consists of an amplified C-band tunable laser, of which output is shaped into a pulse train by a Mach-Zehnder modulator. The pulse duration is 200 ps, at a repetition rate of 25 MHz, and the modulator can be bypassed to probe SHG with CW light. We inject light onto the waveguide fundamental mode with a lensed fiber and evaluate the in-coupling loss to 4.5 dB. At the chip output, a microscope objective collimates the light towards a silicon power detector. In order to extinct the remaining pump light, as well as the visible light from third-harmonic generation (THG), an assembly of short and long pass filters can be placed on the beam path. Out-coupling losses are harder to estimate because of higher scattering at shorter wavelengths. We consider a lower boundary of 5 dB including the attenuation coming from the filter assembly. Alternatively, butt coupling the waveguide output to a multimode fiber allows for the measurement of output spectra.

We first inject the pulsed pump in waveguide (i) with 90 W of coupled peak power on the TM mode, which corresponds to an intensity of about 9 $GW/cm^2$. The weak initial SH, reaching the detector at nanowatt level, increases to approximately 200-250 µW average on the detector (40-50 mW peak) within 25 to 30 minutes. In this interval, we keep the pump power constant. In this configuration, we observe that any SH generated in (i) is TM polarized, like the pump. Fig. 2a shows different SH growth curves as a function of time for pump wavelengths of 1539, 1544 and 1549 nm. A visible camera images the light scattered out-of-plane at the end facet at different growth points. We observe a constant THG (green light) in all growth cases together with an oscillating SH signal (red light). In this waveguide, the growth curve envelopes feature a clear SH power enhancement of at least 30 dB. All the saturation levels are comparable, and slight variations come from non-identical coupling conditions. An example of pump and SH spectrum after growth is shown in Fig. 2b, proving the negligible pump broadening through the waveguide.

We then probe SHG with a CW and tunable laser, of which coupled power is kept constant at 350 mW. Under such light exposure, we do not notice any SHG evolution over time. The SH power is recorded by the photodetector and plotted as a function of wavelength as shown in Fig. 3a. We performed probing after four pumping cases: 1539, 1542, 1544 and 1549 nm. In all cases, one notices a clear peak near the pulsed pump wavelength, indicating that the pump has previously inscribed a modification of the waveguide second-order nonlinearity with the correct periodicity (i.e. a $\chi^{(2)}$ grating). The QPM condition is therefore fulfilled every time we change the pump wavelength, in spite of the waveguide dispersion. The peaks 3 dB bandwidth is about 2.5 nm (at the pump wavelength), while their maxima are located 10 dB higher than what can be considered as the "floor level" for SH. In the 1544 nm pump case, the quadratic relationship between the CW probe power and the SH power is shown in Fig. 3b.

Because of the waveguide birefringence, changing the pump polarization alters the phase mismatch between the pump and the SH. We thus expect the $\chi^{(2)}$ grating period to update when the pump polarization changes. To verify this, we pump waveguide (i) alternatively on the TE and TM mode, keeping the same pump wavelength (1544 nm) and peak power (90 W), and we measure the SH component parallel to the pump polarization thanks to a polarizer at the chip output. We first pump on the TE mode, and observe the TE SH component grow until saturation. We then rotate the polarizer by 90° and observe that the TM SH component is weaker than the TE one by 30 dB. Setting the pump on the TM mode, the TM SH grows again by three orders of magnitude. After another 90° polarizer rotation, only a negligible SH power is detected. However switching the pump back to the TE mode triggers the SH growth again, up to a similar saturation level. Fig. 3c shows the corresponding SH power curves. This experiment is a further demonstration that any $\chi^{(2)}$ grating optically inscribed in the waveguide is overwritten when the coherence length between the pump and SH is changed.

We tested waveguide (ii) by pumping in the TE fundamental mode, a configuration that cancels THG. As explained in the Supplementary Information, we suspect THG to have a detrimental impact on the growth process. We observe again the harmonic growth over time as shown in Fig. 4a for various pump wavelengths and powers. Keeping the coupled pump power constant, we observe similar saturation levels and growth duration when different C-band pump wavelengths are used. Moreover, SHG reaches a saturated average power of 600 μW (120 mW peak) onto the detector when the coupled pump peak power is about 90 W. It corresponds to an estimated SH peak power of 500 mW generated in the waveguide, and to $1.8 \cdot 10^{-6}$ $W^{-1}cm^{-2}$ of efficiency. As before, probing with a CW laser reveals phase matching peaks around the writing wavelengths of 1537 and 1550 nm (see Fig. 4b). In this waveguide, the power levels reached during poling were 280 μW and 600 μW on the detector, at 1550 nm and 1537 nm, respectively, for we used different pump peak powers. The same ratio is thus observed between the SH peaks after the CW probing. Finally, probing the SH power over multiple days indicates no decrease in SHG efficiency (see Fig. 4c), and confirms the long-term and stable grating inscription.

The effect reported here is qualitatively similar to the all-optical SHG enhancement reported in silica fibers pumped by a kW-level pulsed Nd:YAG laser[14]. Researchers regarded the asymmetric emission of photocarriers due to coherent multi-photon absorption as the underlying mechanism[15,16]. The static space-charge field that results from carrier displacement has a periodicity that directly depends on the coherence length between the pump and SH. An effective $\chi^{(2)}$ thus arises as the product of $\chi^{(3)}$ and the built-in field, with the correct periodicity to allow for QPM at the pump wavelength[17]. The gratings are long term inscribed following the trapping of the displaced carriers by deep, localized defect states[16], such as the Si-Si defect states inside the band gap of SiN, extensively studied in the frame of electronic memories[18]. We examine further the physics of the photogalvanic effect in SiN in the Supplementary Information.

Assuming quasi-phase matched SHG with a sine-modulated $\chi^{(2)}$ grating[19], we can estimate the magnitude of $\chi^{(2)}$ in our waveguides from Eq. 1. Here $\omega_{sh}$ is the SH frequency, $P_p$ the coupled pump power, $L$ the waveguide length and $S$ the overlap integral between the pump and SH mode.

$$P_{sh} = \left( \frac{2\omega_{sh} \chi^{(2)} P_p L S}{\pi c n_{sh}} \right)^2 \qquad (1)$$

With the help of a mode solver, we compute the effective indices and the mode overlap of the TM modes at the pump and SH frequencies in waveguide (i). We assume the SH probed at 1544 nm to be on the 8$^{th}$ dipolar mode. From the simulated mode profile and the data from Fig. 3b, we estimate a peak $\chi^{(2)}$ value of about 0.3 pm/V, and we calculate the grating period to be $\Lambda = 2\pi / \left| \beta_{sh} - 2\beta_p \right| \approx 43$ µm (where $\beta$ is the mode propagation constant). For waveguide (ii), the SH is expected to propagate on the 8$^{th}$ TE mode. We retrieve that $\chi^{(2)} \approx 0.15$ pm/V, consistent with the estimation in waveguide (i), and a grating period of about 64 µm at 1537 nm, thanks to the data from Fig. 4a. All the mode simulations and calculations details are presented in the Supplementary Information. Finally, via the relation $\chi^{(2)} =$ $3\chi^{(3)}E_{DC}$[16], we retrieve a space-charge field magnitude of $10^8$ V/m, using a third-order susceptibility of $10^{-21}$ m²/V² [20]. The DC field and $\chi^{(2)}$ magnitude we estimate are comparable to those obtained by applying a voltage across the waveguide structure[20].

In summary, we have experimentally demonstrated an all-optical and reconfigurable SHG enhancement by more than 30 dB in SiN waveguides, after irradiation with a pulsed laser. The enhancement results from persistent inscription of a second-order susceptibility grating in the waveguide. The grating period automatically adapts to a modified coherence length between the pump and the harmonic, allowing for quasi phase matched SHG over the whole C-band.

The presented findings are highly relevant for integrated nonlinear devices. They not only demonstrate a susceptibility of the popular SiN platform to strong irradiation, significantly altering the waveguide properties, but also represent a method to provide QPM for integrated devices without additional complex fabrication process. The overall conversion efficiency can easily be increased using longer waveguides. Moreover, writing gratings in materials with stronger $\chi^{(3)}$ like Si-rich SiN[21] may allow this method to compete with lithium niobate technology.

**References**

1. Levy, J. S., Foster, M. A., Gaeta, A. L. & Lipson, M. Harmonic generation in silicon nitride ring resonators. Opt. Express **19**, 11415-11421 (2011).

2. Logan, D. F. et al. Harnessing second-order optical nonlinearities at interfaces in multilayer silicon-oxy-nitride waveguides. Appl. Phys. Lett. **102**, 061106 (2013).

3. Chang, L., Li, Y., Volet, N., Wang, L. Peters, J. & Bowers, J. E. Thin film wavelength converters for photonic integrated circuits. Optica **3**, 531-535 (2016).

4. Leuthold, J., Koos, C. & Freude, W. Nonlinear silicon photonics. Nature Photonics **4**, 535-544 (2010).

5. Moss, D. J., Morandotti, R., Gaeta, A. L. & Lipson, M. New CMOS-compatible platforms based on silicon nitride and Hydex for nonlinear optics. Nature Photonics **7**, 597-607 (2013).

6. Brasch, V. et al. Photonic chip–based optical frequency comb using soliton Cherenkov radiation, Science **10**, 1126 (2015).

7. Michel, J., Liu, J. & Kimerling, L. C. High-performance Ge-on-Si photodetectors. Nature Photonics **4**, 527 - 534 (2010).

8. Tanzilli, S. et al. Highly efficient photon-pair source using periodically poled lithium niobate waveguide. Electron. Lett. **37**, 26–28 (2001).

9. Bloembergen, N., Chang, R. K., Jha, S. S. & Lee, H. C. Optical second-harmonic generation in reflection from media with inversion symmetry. Phys. Rev. **174**, 813 (1968).

10. Litwin, J. A., Sipe, J. E. & van Driel, H. M. Picosecond and nanosecond second-harmonic generation from centrosymmetric semiconductors. Phys. Rev. B **31**, 5543 (1985).

11. Guyot-Sionnest, P., Chen, W. &. Shen, R. General considerations on optical second-harmonic generation from surfaces and interfaces. Phys. Rev. B **33**, 8254 (1986).

12. Ning, T. et al. Efficient second-harmonic generation in silicon nitride resonant waveguide gratings. Opt. Lett. **37**, 4269-4271 (2012).

13. Pfeiffer, M. H. P. et al. Photonic Damascene process for integrated high-Q microresonator based nonlinear photonics. Optica **3**, 20-25 (2016).

14. Österberg U. & Margulis W. Dye laser pumped by Nd:YAG laser pulses frequency doubled in a glass optical fiber. Opt. Lett. **11**, 516-518 (1986).

15. Anderson, D. Z., Mizrahi, V. & Sipe J. E. Model for second-harmonic generation in glass optical fibers based on asymmetric photoelectron emission from defect sites. Opt. Lett. **16**, 796-798 (1991).

16. Dianov, E. M. & Starodubov, D. S. Photoinduced generation of the second harmonic in centrosymmetric media. Quantum Electronics **25**, 395-407 (1995).

17. Farries, M. C., Russel, P. St.J., Fermann, M. E. & Payne, D. N. Second-harmonic generation in an optical fibre by self-written $\chi^{(2)}$ grating. Electron. Lett. **23**, 322-324 (1987).

18. Gritsenko, V. A., Perevalov, T. V., Orlov, O. M. & Krasnikov, G. Ya. Nature of traps responsible for the memory effect in silicon nitride. Appl. Phys. Lett. **109**, 062904 (2016).

19. Boyd, R. W. *Nonlinear Optics, 3rd ed.* (Academic Press, San Diego, 2008).

20. Puckett, M. W. et al. Observation of second-harmonic generation in silicon nitride waveguides through bulk nonlinearities. Opt. Express **24**, 16923-16933 (2016).

21. Krückel, C. J. et al. Linear and nonlinear characterization of low-stress high-confinement silicon-rich nitride waveguides. Opt. Express **23**, 25827-25837 (2015).

## Methods

### Waveguide fabrication

The waveguide devices were fabricated using the photonic Damascene process approach[13]. The process starts by patterning the waveguides as well as a dense filler pattern into a hardmask of amorphous silicon on a 4 µm thick wet thermal silicon oxide. The structures are then transferred into the preform using a dry etch process based on He and $C_4F_8$. Next the waveguide trenches in the preform are filled with LPCVD silicon nitride, deposited in one run up to the desired thickness. The dense filler pattern efficiently releases film stress and prevents cracking of the SiN thin film. The excess SiN is removed using chemical mechanical polishing, proving a smooth and planar wafer surface. Finally the wafer is annealed to drive out residual hydrogen in the films (1200°C, 24h, $N_2$ atmosphere) and cladded with low temperature oxide (LTO), before being separated into individual chips.

### Acknowledgements

This work was supported by the European Research Council (ERC) under grant agreement ERC-2012- StG 306630-MATISSE, and by contract HR0011-15-C-0055 from the Defense Advanced Research Projects Agency (DARPA), Defense Sciences Office (DSO). SiN waveguide samples were fabricated in the EPFL Center of MicroNanotechnology (CMi).

### Author contributions

A.B. and D.G performed the experiments and carried out the theoretical analysis. M.H.P.P fabricated the SiN waveguides. S.K. contributed to the experiments. T.J.K supervised the fabrication of the waveguides. C.-S.B. supervised experiments in the Photonics Systems Laboratory. All authors contributed to the writing of the manuscript.

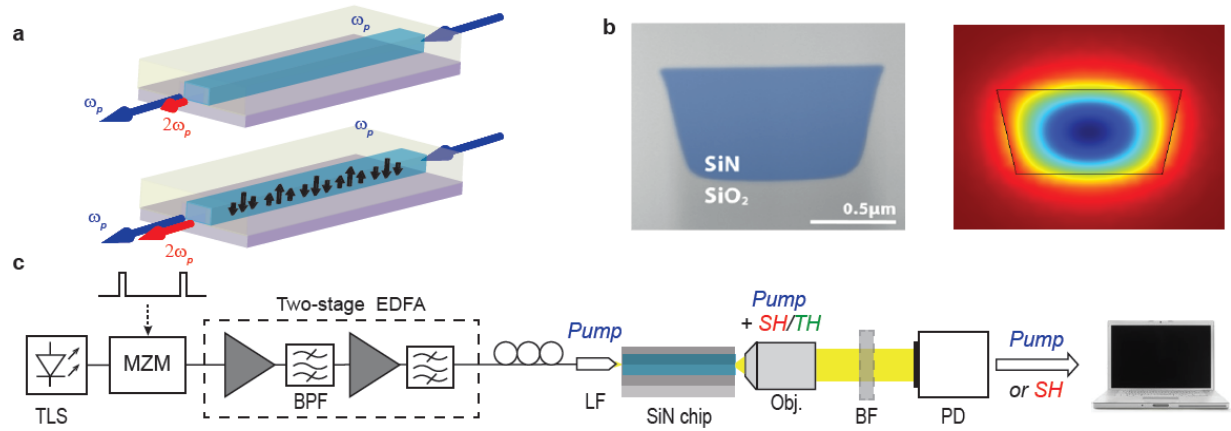

**Figure 1 | Physical principle, waveguide and experimental setup. a,** Illustration of the $\chi^{(2)}$ grating inscription in a SiN waveguide. After irradiation, a spatially periodic DC field builds up and reinforces SHG. **b,** Scanning electron microscope picture of a waveguide cross-section (top) and simulation of the pump TM mode profile (bottom). **c,** Experimental set-up. TLS: tuneable laser source, MZM: Mach-Zehnder modulator, EDFA: erbium-doped fiber amplifier, BPF: fiber Band Pass Filter, LF: lensed fiber, BF: free-space Block Filter assembly, PD: Power Detector.

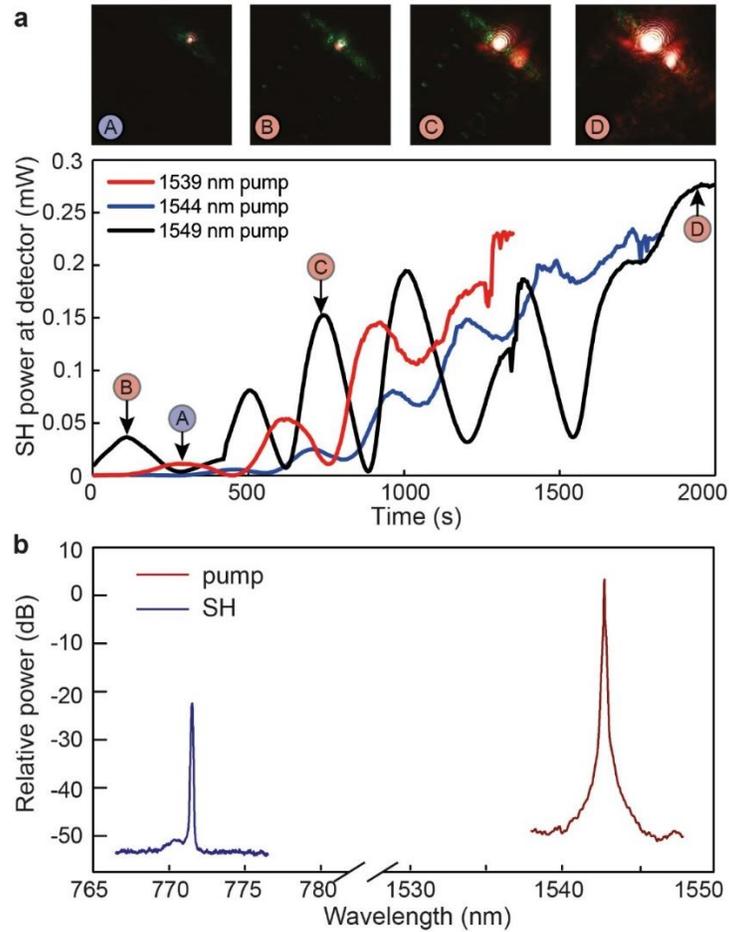

**Figure 2 | Harmonic growth in waveguide (i). a,** Growth curves of the SH average power over time, for pump wavelengths of 1539, 1544 and 1549 nm. The coupled peak power is 90 W. The top pictures show the light scattered at the end facet, coming either from SHG (red) or THG (green). The labels (A)-(D) indicate the instant and for which pumping wavelength the picture was taken. **b,** Pump and SH spectra at the waveguide output for a coupled peak power of 90 W, after the SH growth. No significant pump broadening is observed.

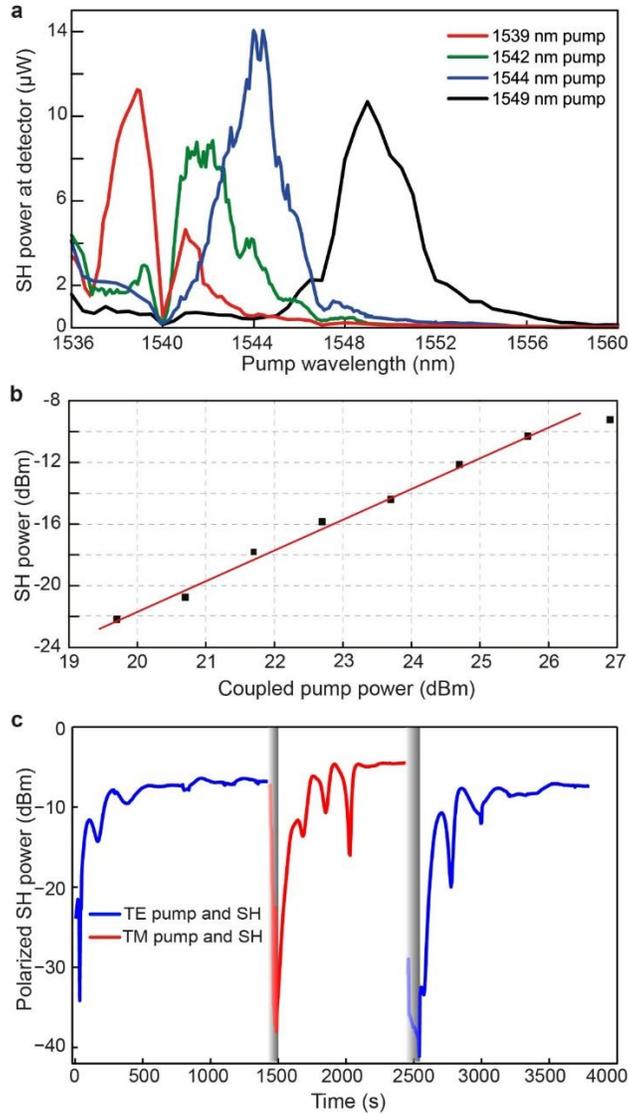

**Figure 3 | Quasi phase matching evidence in waveguide (i). a,** SH power as a function of the CW probe wavelength, in waveguides previously pumped at 1539, 1542, 1544 and 1549 nm. Phase-matching peaks are observed around each pump wavelength. **b,** SH power (estimated in the waveguide) as a function of the coupled CW probe power. The probe is centered at 1544 nm, and the grating was previously inscribed at the same wavelength. The squares are experimental points while the red line is a just a linear fit with a slope of 2. The observed saturation at high power comes from coupling instabilities. **c,** Power of the TE or TM SH component over time when the pump polarization is switched from the TE to TM mode, and back to TE. The grey rectangles represent switching points. The polarizer is first rotated by 90°, entailing a measured SH power drop. The pump polarization is then aligned parallel to the polarizer axis, triggering the SH component growth.

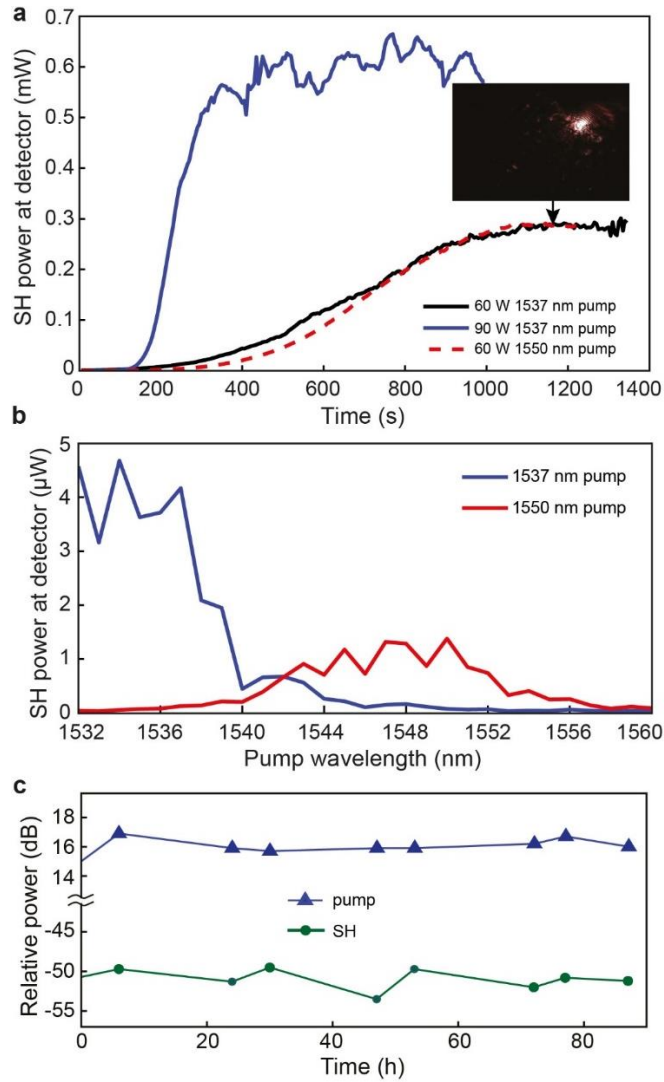

**Figure 4 | Grating dynamics in waveguide (ii). a,** SH growth over time, for pump wavelengths of 1537 and 1550 nm. The coupled peak power is varied between 60 and 90 W. The inset shows the light scattered at the end facet once saturation is reached, with very little green emission. b, SH power as a function of the CW probe wavelength, in a waveguide previously pumped at either 1537 nm or 1550 nm. **c,** Persistence measurement showing the CW probe and SH power over more than 80 h of operation. Both quantities are measured at the waveguide output.

# Supplementary Information

**Coherent photogalvanic effect in silicon nitride**

The photogalvanic effect corresponds to the apparition of a current in a medium (either centrosymmetric or not) under uniform illumination. In fibers, researchers postulated that asymmetric photoemission takes place from polar defect centers (such as GeO$_2$-related defects in silica) under illumination by a pump and its frequency-doubled counterpart, as a result from the interference between three-photon absorption and both the two- and four-photon absorption[1]. The electron ejection occurs along the defect site dipole axis, but since SiO$_2$ and SiN are amorphous, the material does not induce any preferential ejection direction. Consequently, the effect does not depend on the pump laser polarization[2]. In a one-dimensional approximation, the algebraic coherent photogalvanic current $j_{ph}$ has a dependence given by the equation below[1].

$$j_{ph} \propto i * \eta_1 |E_p|^2 |E_{sh}| \left( \eta_2 |E_p|^4 - |E_{sh}|^2 \right) \exp\left( i(\Phi_{sh} - 2\Phi_p) \right) + \text{c.c.} \qquad (1)$$

Where $E$ is the local electric field of the pump or SH (denoted by the subscript $p$ or $sh$, respectively), $\Phi$ is the phase of the corresponding fields (same subscript), $\eta_1$ and $\eta_2$ are coefficients of multi-photon absorption, and c.c. indicates the complex conjugate. The exponential term denotes the coherent nature of the process, and leads to spatially periodic charge separation. Charges therefore migrate in a preferential direction and $j_{ph}$ gets associated to a space-charge DC field $E_{DC}$. The relation between the two is given by the photoconductivity $\sigma$, proportional to the total density of carriers promoted to the conduction band. This relation is $j_{ph} = \frac{E_{DC}}{\sigma}$. For a certain level of displaced carriers, $E_{DC}$ reaches a threshold value that prevents a further migration of photo-carriers, and the process saturates. Finally the space-charge field persistence is explained by the trapping of displaced carriers by deep defect states near the conduction band.

In the case of SiN, Si-Si bonds appear to act as traps for both holes and electrons, located 1.4 eV away from the valence and conduction bands, inside the 4.6 eV bandgap[3]. It means that electron promotion from an occupied defect state to the conduction band requires 3.2 eV. The photon energy for a 1.55 µm pump is 0.8 eV, while the SH photon energy is 1.6 eV. The interferential multi-photon absorption process described in Eq. 1 considers absorption processes exactly corresponding to a total energy of 3.2 eV for both terms. It shows that these Si-Si defects are good candidates to explain asymmetric carrier ejection, without even considering intermediate state absorption[2]. Moreover, SiO$_2$-SiN structures can localize electrons and holes for multi-year lifetimes[4]. Since these Si-Si defects also entail localized states 1.4 eV below the conduction band, they can potentially trap free carriers as well. It would result in a long-lived and thermalization-immune localization of the photo-carriers, in agreement with the observed grating persistence.

In our experiments, we have also observed the adverse effect of green light from THG on the SH enhancement dynamics, manifesting itself for instance by oscillations in the growth curves in waveguide (i). Moreover, we observed the fast erasure of $\chi^{(2)}$ gratings in case THG is strong enough. In fibers, visible light is able to re-ionize the trapped electrons that entail the space charge field[5]. In our case, green light has an energy of 2.4 eV, sufficient to excite the carriers trapped in defect states near the conduction band and lead to their recombination, and to the grating erasure.

**Evaluation of $\chi^{(2)}$ after grating inscirption**

The overlap integral *S* between the pump and SH mode, appearing in Eq. 1 from the main text, has an analytical formula derived from the coupled mode theory[6] and given by the equation below.

$$S = \iint_A E_p^2(x,y) E_{sh}(x,y) \, \text{sign}(\chi^{(2)}) \, dxy \tag{2}$$

In this formula, the integration is carried out over the entire waveguide cross section. The electric fields are normalized to the power flux ($P_z$) across the waveguide cross section given by the mode solver:

$$\iint_A |E_i|^2 dxy = \frac{2P_z}{c\varepsilon_0 n},$$ where $i=x$ or $y$, depending on the field polarization.

The local value of the space charge field $E_{DC}$ has an orientation and a magnitude that depends upon the phase difference between the pump and SH. The $\chi^{(2)}$ grating, which originates from this field via the third-order susceptibility tensor ($\chi^{(2)} = 3\chi^{(3)} E_{DC}$)², also follows this rule. In fact, this $\chi^{(2)}$ dependence holds whatever the microscopic model behind the phenomenon, as verified in the main text by measuring the QPM peaks. Yet, in our waveguides, the SH propagates on a dipolar or tripolar mode, of which lobes are in quadrature. The pump field, in the fundamental mode, has a constant phase over the waveguide cross section. The space charge field $E_{DC}$ must therefore have opposite signs in each lobe region. Consequent, $\chi^{(2)}$ also switches sign from a lobe to the other, as does the SH electric field phasor. The overlap integral can thus be rewritten as $S = \iint_A E_p^2(x,y) |E_{sh}(x,y)| dxy$.

We performed the evaluation of $\chi^{(2)}$ and of the grating period given in the main text by computing the waveguide dispersion and mode profile (see Supplementary Fig. 1a and 1d) with a finite-element solver (COMSOL Multiphysics). In all cases, the SH is assumed to propagate on the higher order mode that yields the smallest index difference (i.e. which yields the best natural phase matching) with the pump. QPM then establishes via the photogalvanic effect to compensate for the index mismatch. In waveguide (i), only the TM modes are considered, since both the pump and harmonic are TM in most of the reported experiments. The SH mode featuring the smallest index difference with the pump is the 8$^{th}$ TM mode, of which profile is shown in Supplementary Fig. 1b. The TM 10$^{th}$ order mode (seen in Supplementary Fig. 1d) is slightly more phase-mismatched. In the case of waveguide (ii), only the TE modes are considered and their index curves are displayed in Supplementary Fig. 1d. The pump propagates over the fundamental TE mode (see Supplementary Fig. 1e), while the second harmonic is in the 8$^{th}$ TE mode, of which profile is reported in Supplementary Fig. 1f.

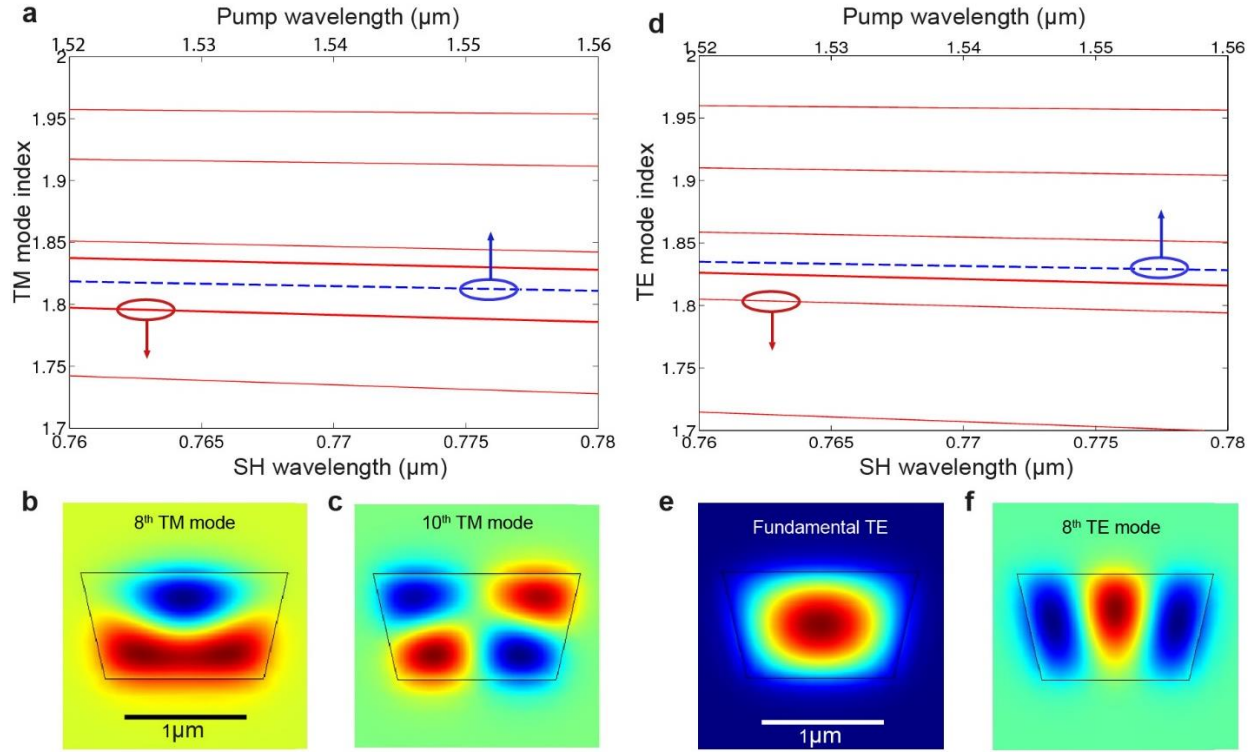

**Supplementary Figure 1 | Mode solving.** . Effective index curves for the TM modes in waveguide (i), as computed with a mode solver. The solid red curves correspond to higher order modes at the SH wavelength, while the dashed blue curve corresponds to the fundamental mode at the pump wavelength. **b,** Simulated profile for the 8$^{th}$ TM mode at 770 nm in (i). This mode corresponds to the curve right over the fundamental curve. **c,** Same for the 10$^{th}$ TM mode in (i), corresponding to the curve right below the fundamental curve. **d,** Effective index curves for the TE modes in waveguide (ii). **e,** Simulated profile for the fundamental TE mode at 1540 nm in (ii). **f,** Simulated profile for the 8$^{th}$ TE mode at 770 nm in (ii). This mode corresponds to the curve right below the fundamental curve.

**References**

1. Anderson, D. Z., Mizrahi, V. & Sipe J. E. Model for second-harmonic generation in glass optical fibers based on asymmetric photoelectron emission from defect sites. Opt. Lett. **16**, 796-798 (1991).

2. Dianov, E. M. & Starodubov, D. S. Photoinduced generation of the second harmonic in centrosymmetric media. Quantum Electronics 25, 395-407 (1995).

3. Gritsenko, V. A., Perevalov, T. V., Orlov, O. M. & Krasnikov, G. Ya. Nature of traps responsible for the memory effect in silicon nitride. Appl. Phys. Lett. **109**, 062904 (2016).

4. Gritsenko, V. A. et al. Excess silicon at the silicon nitride/thermal oxide interface in oxide–nitride–oxide structures. J. Appl. Phys. **86**, 3234 (1999).

5. Ouellette, F., Hill, K. O. & Johnson, D. C. Light-induced erasure of self-organized $\chi^{(2)}$ gratings in optical fibers. Opt. Lett. **13**, 515-517 (1988).

6. Yariv, A. Coupled-mode theory for guided wave optics. IEEE J. Quant. Electron. **9**, 919-933 (1973).